\documentclass[structabstract]{aa}  
\usepackage{graphicx}
\usepackage{txfonts}
\usepackage{natbib,twoopt}
\begin{document}
\title{Kinematic imprint of clumpy disk formation on halo objects}


\author{Shigeki Inoue
  \inst{1,2}
}

\institute{Mullard Space Science Laboratory, University College London, Holmbury St. Mary, Dorking, Surrey, RH5 6NT, UK\\
  \email{shigeki.inoue@nao.ac.jp}
  \and
  National Astronomical Observatory of Japan, Mitaka, Tokyo 181-8588, Japan
}

\date{Received September 15, 1996; accepted March 16, 1997}


\abstract
    {Clumpy disk galaxies in the distant universe, at redshift of $z\hspace{0.3em}\raisebox{0.4ex}{$>$}\hspace{-0.75em}\raisebox{-.7ex}{$\sim$}\hspace{0.3em}1$, have been observed to host several giant clumps in their disks. They are thought to correspond to early formative stages of disk galaxies. On the other hand, halo objects, such as old globular clusters and halo stars, are likely to consist of the oldest stars in a galaxy (age $\hspace{0.3em}\raisebox{0.4ex}{$>$}\hspace{-0.75em}\raisebox{-.7ex}{$\sim$}\hspace{0.3em}10~{\rm Gyr}$), clumpy disk formation can thus be presumed to take place in a pre-existing halo system.}
    {Giant clumps orbit in the same direction in a premature disk and are so massive that they may be expected to interact gravitationally with halo objects and exercise influence on the kinematic state of the halo. Accordingly, I scrutinize the possibility that the clumps leave a kinematic imprint of the clumpy disk formation on a halo system.}
    {I perform a restricted $N$-body calculation with a toy model to study the kinematic influence on a halo by orbital motions of clumps and the dependence of the results on masses (mass loss), number, and orbital radii of the clumps.}
    {I show that halo objects can catch clump motions and acquire disky rotation in a dynamical friction time scale of the clumps, $\sim0.5~{\rm Gyr}$. The influence of clumps is limited within a region around the disk, while the halo system shows vertical gradients of net rotation velocity and orbital eccentricity. The significance of the kinematic influence strongly depends on the clump masses; the lower limit of postulated clump mass would be $\sim5\times10^8~{\rm M_\odot}$. The result also depends on whether the clumps are subjected to rapid mass loss or not, which is an open question under debate in recent studies. The existence of such massive clumps is not unrealistic. I therefore suggest that the imprints of past clumpy disk formation could remain in current galactic halos.}
    {}
    
    \maketitle
%

\section{Introduction}

Although disk galaxies in the local universe generally have relatively smooth stellar distributions in which small star clusters of $M_{cl}\hspace{0.3em}\raisebox{0.4ex}{$<$}\hspace{-0.75em}\raisebox{-.7ex}{$\sim$}\hspace{0.3em}10^3~{\rm M_\odot}$ reside \citep[e.g.,][]{ll:03}, some galaxies in the high redshift universe (redshift of $z\sim$ 1 -- 5) have been observed to have clumpy structures in which massive \textit{clumps} are forming \citep[e.g.,][]{bar:96,eeh:04,eer:07,eef:09,eem:09,ee:05,fgl:06,fgb:09,fsg:11,g:06,g:08,gnj:11,t:10,p:10,ggf:11,spc:11,sss:12,nsg:12,w:12,l:12}. Regardless of their irregular morphologies, the galaxies indicate clear rotation in spectroscopic observations as signatures of disk structures, although some considerable fraction of them may be ongoing mergers \citep[e.g.,][]{w:06,fgb:09,p:10}. These are referred to as clump clusters and chain galaxies, depending on viewing angles seen face-on or edge-on, respectively. In such galaxies, a single clump was observed to have, at the largest, a mass of $M_{cl}\sim10^9-10^{10}~{\rm M_\odot}$. Recently, \citet{hcw:12} also found a rotating clumpy gas disk in a sub-millimeter galaxy at a redshift $z\simeq4$. The clumpy galaxies are thought to be disk galaxies in early stages of their formation. Numerical simulations showed that dynamical instability due to the gas-rich nature of a premature disk can lead to clump formation: the clumps fall into the galactic center by dynamical friction and may form a central bulge. Such numerical studies demonstrated that a clumpy galaxy can finally evolve into a disk galaxy \citep{n:98,n:99,isg:04,isw:04,bee:07,ebe:08,ebe:08b,be:09,atm:09,dsc:09,abj:10,cdb:10,cdm:11,is:11,is:12}.

However, these results do not necessarily mean that all of the current disk galaxies were once clumpy galaxies. \citet{ee:05} have observationally discussed that clumpy galaxies have too high a surface density to be current disk galaxies such as the Milky Way and that they would evolve into other types of galaxy (e.g., elliptical galaxies by mergers). Furthermore, \citet{mgc:12} performed simulations of modeled galaxies joined to a cosmological run and demonstrated that clumpy disk formation is preceded by intense gas accretion at high redshift and that clumpy disk formation would not be common in isolated environments. 

How common was clumpy disk formation in the universe? Can we know which disk galaxy was once a clumpy galaxy? To answer these questions, it is important to seek a clue concerning the past clumpy disk formation in current galactic structures. In this paper, I focus on halo objects such as old globular clusters (GCs) and halo stars (HSs), which are likely to consist of the oldest stars in a galaxy. Generally, GCs are classified into metal-poor and -rich ones. Metal-poor GCs are usually old (age $\hspace{0.3em}\raisebox{0.4ex}{$>$}\hspace{-0.75em}\raisebox{-.7ex}{$\sim$}\hspace{0.3em}10~{\rm Gyr}$), although there are some exceptions and sub-groups in their classification \citep[e.g.,][and references therein]{hr:79,h:91,h:01,bs:06}.

Clumpy disk galaxies, on the other hand, can be observed even at a redshift of $z\sim1$ ($\sim8~{\rm Gyr}$ in look-back time). Hence, clumpy disk formation can be thought to postdate the formation epoch of halo objects and expected to take place in a \textit{pre-existing} halo if the halo objects have an internal origin, such as the collapse of proto-galactic gas\footnote{As another scenario, halo objects may be brought in by accretion events of dwarf galaxies after the clumpy phase \citep[e.g.,][]{ans:06}. Composite models are also possible. \citet{mg:04} have estimated that nearly half of the mass of the Galactic stellar halo is from external accretion and the other half is from an internal origin.}, which was the concept of \citet{els:62}. If clumps are many and massive enough to interact gravitationally with halo objects, orbital motions of the clumps may exercise influence on the kinematic state of the halo objects. \citet{ebe:08} and \citet{is:11} have conducted numerical simulations that showed that a density profile of a dark matter (DM) halo can be changed due to kinematic heating by giant clumps. Their results also imply that kinematics of halo objects may be affected by the clumps.

In this paper, I perform restricted $N$-body calculations using simple toy models, demonstrate how giant clumps exercise influence on halo objects, and examine dependence on masses, number, and orbital radii of the clumps. Additionally, I discuss the impact of mass loss of the clumps on my result, which have been recently suggested to be caused by a strong gas outflow. Although, in my study I discuss the kinematics of GC systems, this can also be applied to HS. I explain my model settings and method in \S\ref{model} and show the results and discuss the kinematic imprint of clumps in \S\ref{result}. I present the discussion and summary of this paper in \S\ref{dis} and \S\ref{sum}, respectively.
\label{sum}

\section{Restricted $N$-body calculation}
\label{model}
As mentioned above, I perform restricted $N$-body calculations with a simple model. I assume a DM halo represented by a rigid potential of a Navarro-Frenk-White model \citep[hereafter NFW]{nfw:97} with a virial mass of $1.0\times10^{12}~{\rm M_\odot}$, a virial radius of $200~{\rm kpc}$, and a concentration parameter of $15$. Indeed, clumpy galaxies at a redshift of $z\sim2$ are thought to live in DM halos of $\sim10^{12}~{\rm M_\odot}$ in which cold stream is expected to take place \citep[e.g.,][]{dbe:09}. 

Initially GCs (and HSs) are assumed to have a distribution with a spherical number density profile of
\begin{equation}
  n_{\rm GC}(r)=n_0\frac{{r_0}^{3.5}}{r^2(r_0+r)^{1.5}},
    \label{density}
\end{equation}
according to observations of GCs in the Milky Way \citep{h:01}, where I set $r_0=4~{\rm kpc}$ and $n_0$ is a constant normalized by the total number of GCs (see below). I assume a truncation on the distribution outside $100~{\rm kpc}$ and a constant density inside $100~{\rm pc}$. The GC system has an equilibrial and isotropic kinematic state determined by a solution of Jeans equations. A single GC is represented by a single mass-less particle.

As a fiducial model, the structure of a clump is represented by a Plummer model with a mass of $M_{\rm cl}=5\times10^8~{\rm M_{\odot}}$ and a Plummer radius of $\epsilon=200~{\rm pc}$. \citet{ee:05} have observed that the numbers of clumps are 5 -- 14 and the averaged clump masses range from 0.21 -- 1.36 $\times10^9~{\rm M_{\odot}}$  in their sample galaxies. In accordance with their observation, my initial condition sets ten clumps in circular orbits at radii of $1, 2, 3, \cdots, 10~{\rm kpc}$ from the halo center. All clumps orbit in the same direction in the same orbital plane. The initial orbital phase is randomly determined on each clump. Additionally, I perform other runs with different settings of the clump mass, number, and the initial radii (see \S\ref{result}). 

Gravity calculation takes into account gravitational forces from the clumps to the GCs and mutual interactions between the clumps. I set the total number of GCs to 100,000. Of course, there is not such a large number of GCs in any given galaxy. However, results of this calculation are independent of the number of GCs since they are mass-less particles. Hence, I employ as large a number of GCs as possible for the sake of good statistics. Dynamical friction is ignored on the GCs, but embodied with the Chandrasekhar formula on the clumps, adopting $\ln \Lambda=8.4$ \citep[][]{c:43,bt:08}. Since a proper value of $\ln\Lambda$ is unclear \cite[e.g.,][]{hfm:03,ffm:06,i:09,i:11}, I set the value to match infall time scale of the clumps with the results of other numerical simulations, $\sim0.5~{\rm Gyr}$ \citep[e.g.,][]{bee:07,ebe:08,cdm:11}. When a distance between two clumps becomes shorter than $\epsilon$, they merge into one, the mass of which is the sum of the two. I conduct the orbital calculation with the second-order leap-frog time integrator and a shared time step of $\Delta t=0.03\times\sqrt{\epsilon^3/(GM_{\rm cl})}=0.057~{\rm Myr}$ until $t=5~{\rm Gyr}$. 

\section{Results}
\label{result}
\subsection{The fiducial model}
\label{fiducial}
In the fiducial model, I set ten clumps in the DM halo; the masses of the clumps are $M_{\rm cl}=5\times10^8~{\rm M_{\odot}}$, following the observation of \citet{ee:05}. Since their galaxy samples have disk diameters of 13 -- 29 ${\rm kpc}$, I set the initial clump orbital radii to 1 -- 10 ${\rm kpc}$ with constant separations. I assume that the clump masses are constant unless they merge.

\begin{figure}
  \includegraphics[width=\hsize]{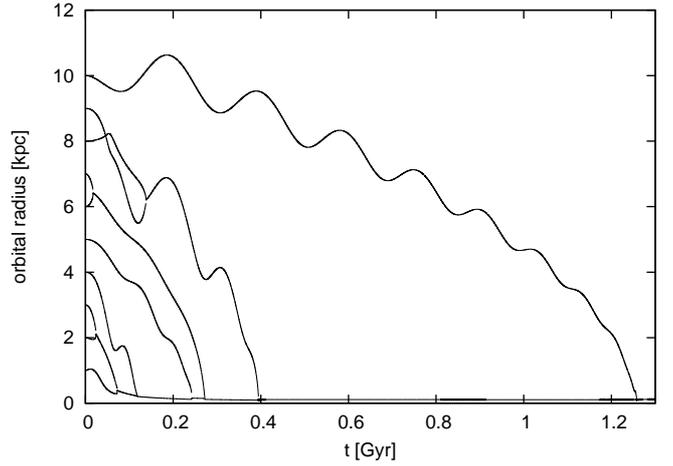}
  \caption{Time evolution of orbital radii of the clumps in the fiducial model.}
  \label{GCs}
\end{figure}
The clumps fall into the galactic center by dynamical friction and finally form a clump-origin bulge. Figure \ref{GCs} shows time evolution of orbital radii of the clumps in the fiducial case. Nine of the ten clumps reach the galactic center by $t\simeq0.4~{\rm Gyr}$ while merging with one another. Although the outermost clump does not experience any merger, it spirals into the center at $t\simeq1.2~{\rm Gyr}$.

\begin{figure}
  \includegraphics[width=\hsize]{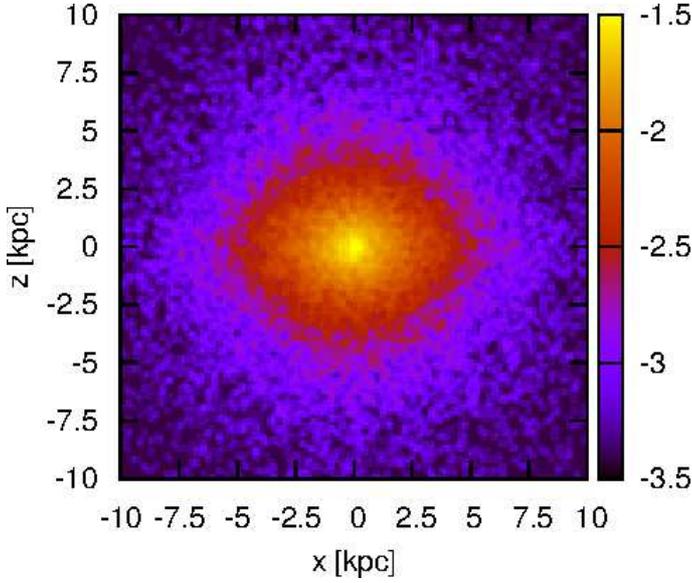}
  \caption{Probability density distribution of the GCs from the edge-on view in the unit of ${\rm kpc^{-2}}$ in the end state of the fiducial calculation. The initial distribution was set to the spherical one, according to Eq. \ref{density}. The probability density is defined as the number of GCs in each bin divided by the total number.}
  \label{dist}
\end{figure}
\begin{figure}
  \includegraphics[width=\hsize]{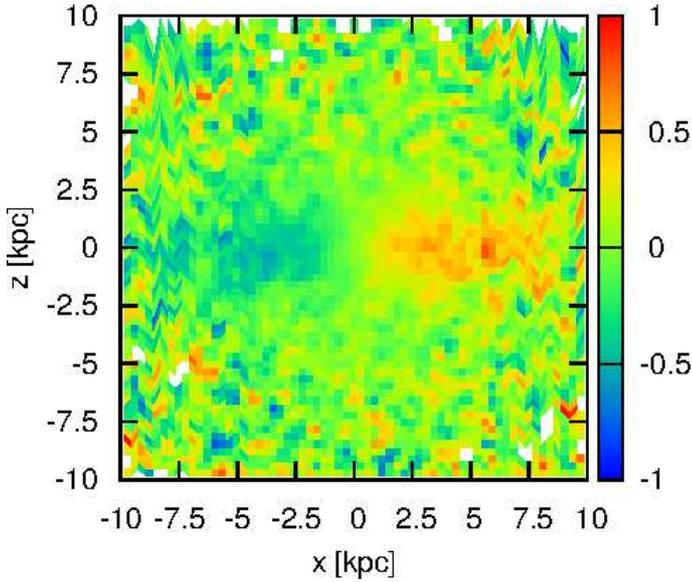}
  \caption{Map of mean LoS velocity from the edge-on view in the fiducial run. The values are normalized by the LoS velocity dispersion inside $r_0=4~{\rm kpc}$, which is $106~{\rm km~s^{-1}}$.}
  \label{los}
\end{figure}
Figure \ref{dist} displays a GC distribution seen edge-on in the end state of the calculation. The figure shows a slightly aspherical distribution, though it was initially assumed to be spherical. Figure \ref{los} shows a line-of-sight (LoS) velocity map. The halo system indicates a clear rotation around the disk plane in the end state, implying that the GCs catch the clump orbital motions through gravitational interaction. 

\begin{figure}
  \includegraphics[width=\hsize]{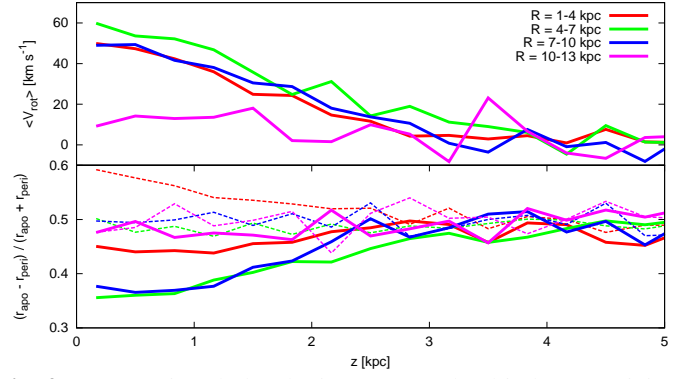}
  \caption{Mean azimuthal velocity (top) and orbital eccentricity, $e\equiv(r_{apo}-r_{peri})/(r_{apo}+r_{peri})$, (bottom) of the GCs as functions of distance from the clump orbital plane in the end state of the fiducial run, where $r_{apo}$ and $r_{peri}$ are apo- and peri-center distances, respectively. Each line indicates a radial range in cylindrical coordinate, $R$. In the bottom panel, GCs rotating prograde and retrograde are separately plotted; the thick and thin lines correspond to prograde and retrograde ones, respectively.}
  \label{dist_z}
\end{figure}
Vertical profiles of mean azimuthal velocities and orbital eccentricities in different radial ranges are shown in Figure \ref{dist_z}. The kinematic influence by the clumps can be seen up to $R\hspace{0.3em}\raisebox{0.4ex}{$<$}\hspace{-0.75em}\raisebox{-.7ex}{$\sim$}\hspace{0.3em}10~{\rm kpc}$ and $z\hspace{0.3em}\raisebox{0.4ex}{$<$}\hspace{-0.75em}\raisebox{-.7ex}{$\sim$}\hspace{0.3em}3~{\rm kpc}$, showing vertical gradients of the rotation velocities and the orbital eccentricities. In this region, the halo indicates slow but clear net rotation and GCs rotating prograde are circularized. In the innermost region, $R=1 - 4~{\rm kpc}$, orbits of retrograde GCs are radialized rather than circularized, and prograde GCs are not strongly circularized. This is probably because of a potential of the massive bulge built up from the clumps. 

In the fiducial run, I see that clump motions can rotate the halo objects if they are many and massive enough. However, some parameters assumed in the calculation are unclear in current observations and clumpy galaxies have been observed to have a variety of their properties. In the following subsections, I perform further calculations with different settings.

\subsection{Fewer clumps}
\label{5clumps}
Although \citet{ee:05} have observed that the average number of clumps is nearly ten in a clumpy galaxy, this may not be a typical number. For example, \citet{fsg:11} and \citet{ggf:11} observed that the median numbers of clumps are five and four in their samples, respectively. Accordingly, I run similar calculations with fewer clumps.

\begin{figure}
  \includegraphics[width=\hsize]{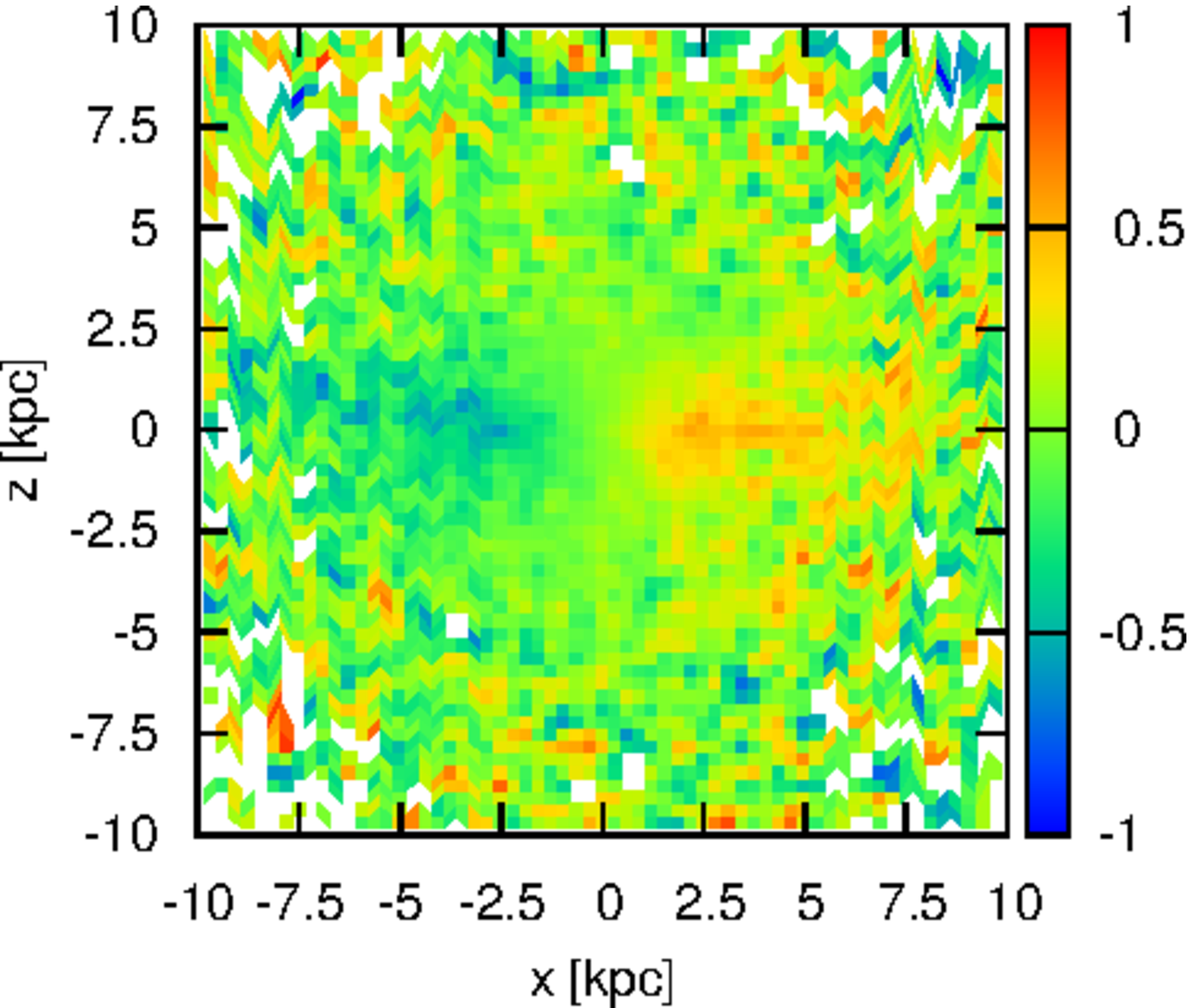}
  \includegraphics[width=\hsize]{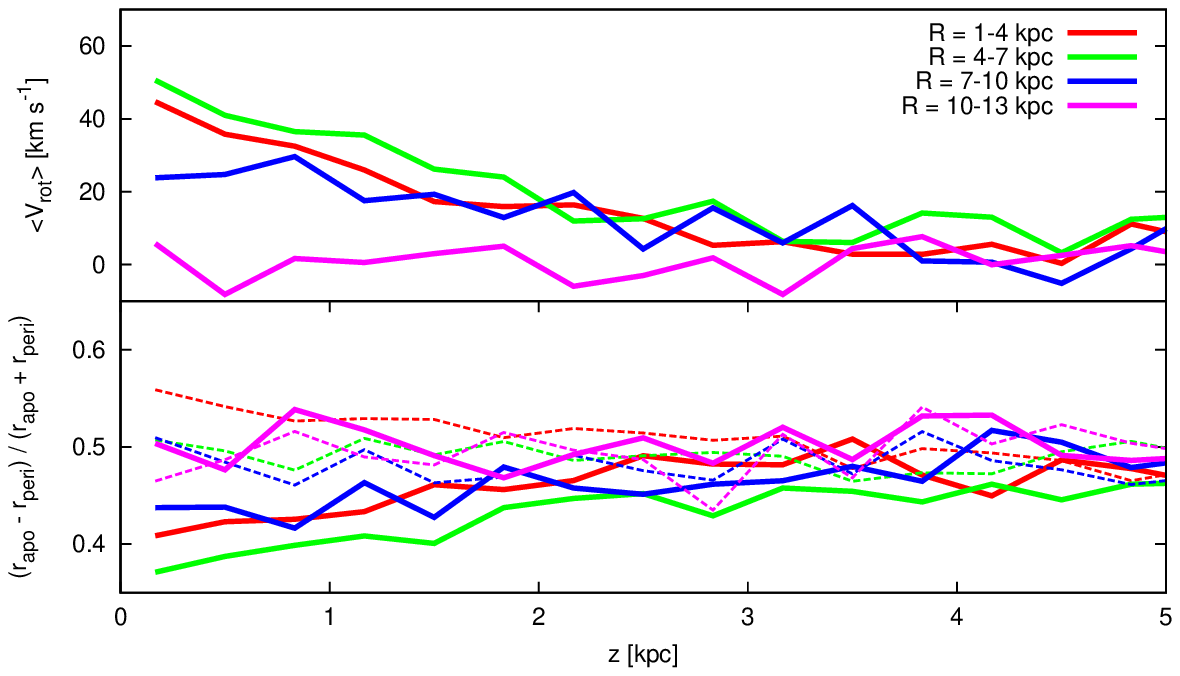}
  \caption{Same as Figure \ref{los} and Figure \ref{dist_z}, but setting five clumps with constant separations of $2~{\rm kpc}$ in the initial state. The LoS velocity dispersion inside $r_0$ is $92.2~{\rm km~s^{-1}}$ in the top panel.}
  \label{wide_5clumps}
\end{figure}
I set five clumps in the halo model, of which initial orbital radii are $2, 4, \cdots, 10~{\rm kpc}$. The other parameters are the same as the fiducial run. Figure \ref{wide_5clumps} displays a LoS velocity map and vertical profiles of mean azimuthal velocities and orbital eccentricities. Although the rotating signature is a little weaker than the fiducial run, significant differences cannot be seen. Hence, the number of clumps does not seem to change the results in the range of 5 -- 10 clumps.

The orbital radii of clumps in the fiducial run may also not be typical. Although giant clumps seem to fall toward the center from $\hspace{0.3em}\raisebox{0.4ex}{$>$}\hspace{-0.75em}\raisebox{-.7ex}{$\sim$}\hspace{0.3em}10~{\rm kpc}$ in cosmological simulations \citep{cdm:11}, \citet{eer:07} had observed that root mean squares of clump positions in galaxies are 1 -- 8 $\rm kpc$. Thus, some clumpy galaxies may be more compact than the fiducial model. Accordingly, I conduct a run setting five clumps at $1, 2, \cdots, 5~{\rm kpc}$ from the galactic center.

\begin{figure}
  \includegraphics[width=\hsize]{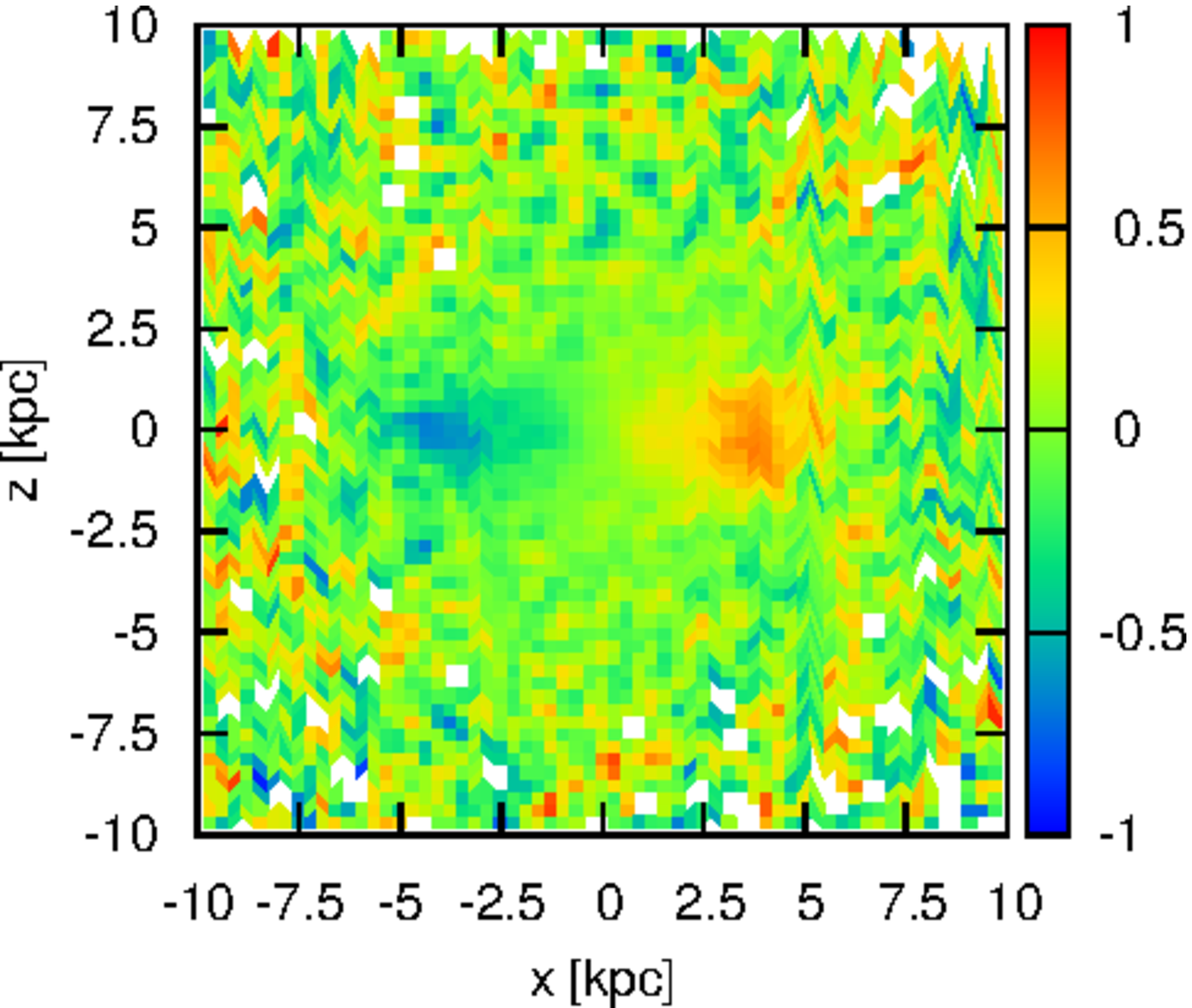}
  \includegraphics[width=\hsize]{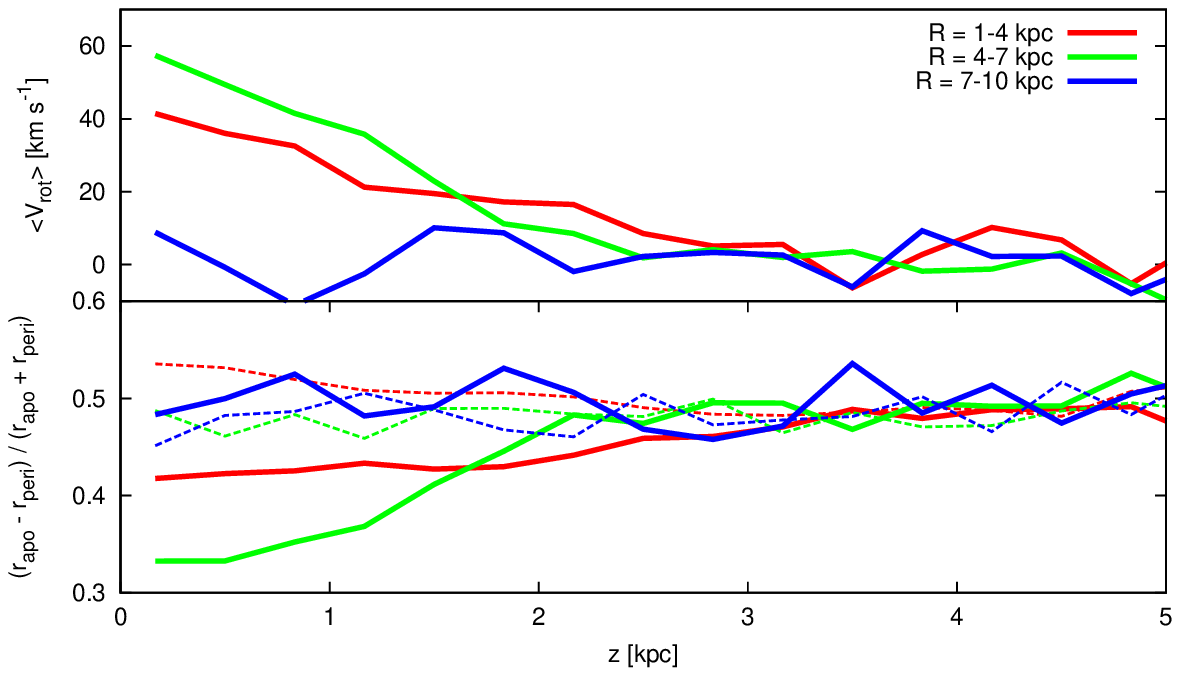}
  \caption{Same as Figure \ref{los} and Figure \ref{dist_z}, but setting five clumps with constant separations of $1~{\rm kpc}$ between $R=1 - 5~{\rm kpc}$ in the initial state. The LoS velocity dispersion inside $r_0$ is $92.5~{\rm km~s^{-1}}$ in the top panel.}
  \label{inner_5clumps}
\end{figure}
The results are shown in Figure \ref{inner_5clumps}. The figure indicates significant rotation and orbital circularization in $R=1 - 7~{\rm kpc}$ as a kinematic imprint of the clump motions. The region of $R>7~{\rm kpc}$ remains intact because of the absence of outer clumps. From these results, I suggest that inner regions of a disk galaxy may retain the disky rotation caused by clump motions since clumps have to pass through the inner regions of galaxies to reach the center, the inner regions therefore cannot avoid the kinematic influence by the clumps, in cases where the clumps are long lived and can reach the galactic center by dynamical friction.

\subsection{Lower mass clumps}
\label{lowmass}
Masses of giant clumps have been observed to range widely in $10^5 - 10^{10}~{\rm M_{\odot}}$ \citep[e.g.,][]{eef:09,eem:09,ggf:11,l:12}. \citet{sss:12} have recently discussed that the most massive clump mass in a clumpy galaxy would be $10^8 - 10^9~{\rm M_{\odot}}$ and present observations have shown that massive clumps reaching $\hspace{0.3em}\raisebox{0.4ex}{$>$}\hspace{-0.75em}\raisebox{-.7ex}{$\sim$}\hspace{0.3em}10^{9}~{\rm M_{\odot}}$ do exist. However, such massive clumps may be rare. Accordingly, I run another simulation setting ten clumps, the masses of which are $M_{cl}=1\times10^8~{\rm M_{\odot}}$. The other parameters are the same as the fiducial run.

\begin{figure}
  \includegraphics[width=\hsize]{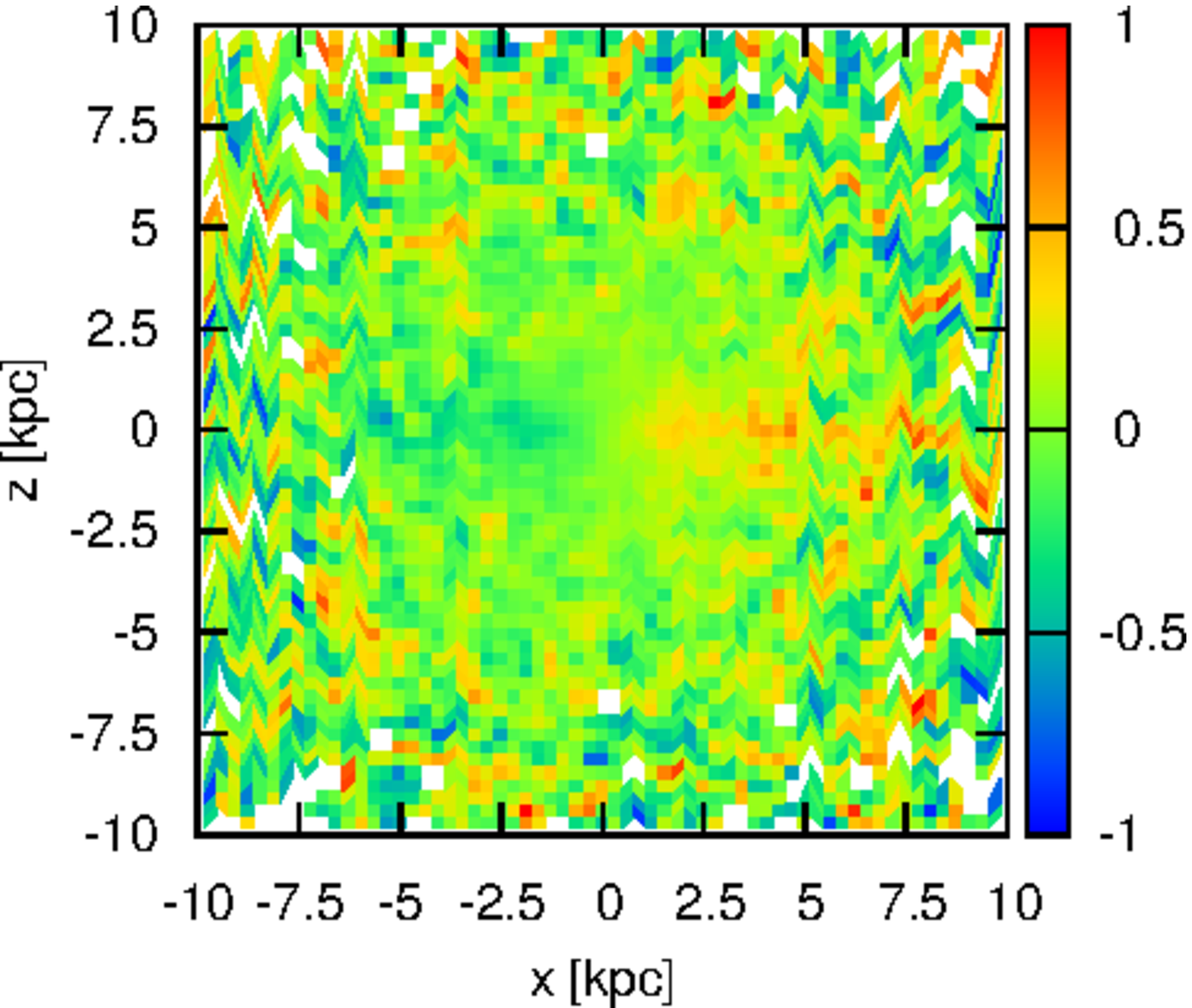}
  \includegraphics[width=\hsize]{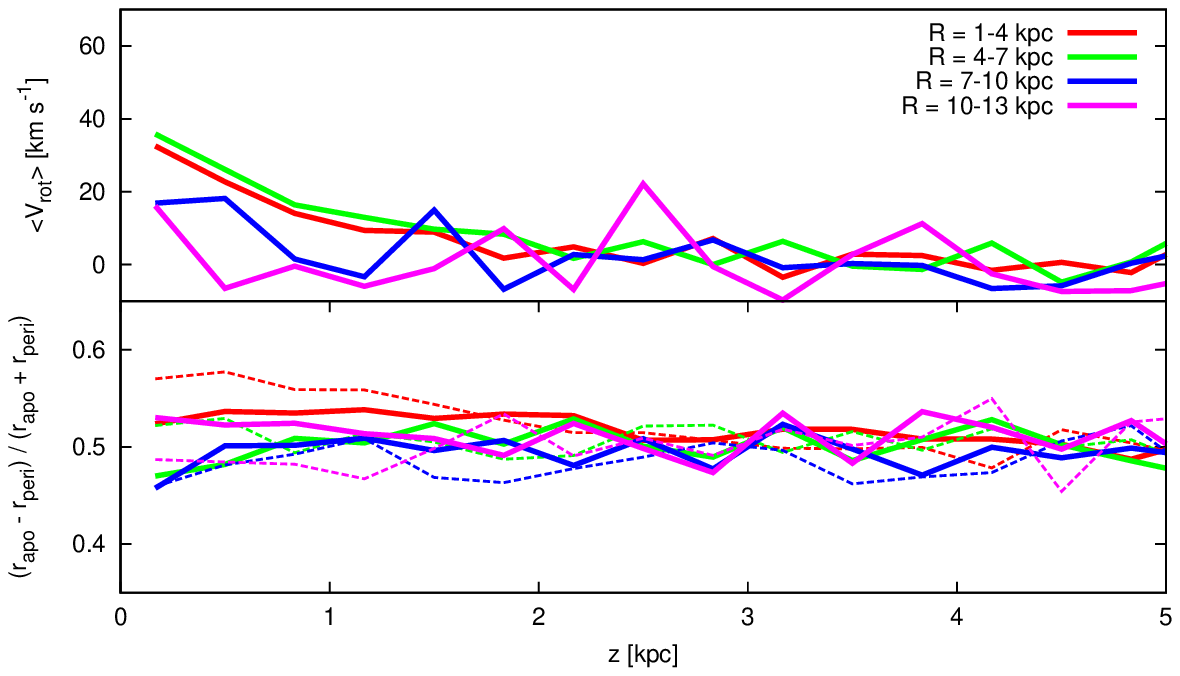}
  \caption{Same as Figure \ref{los} and Figure \ref{dist_z}, but ten clumps of $1\times10^8~{\rm M_{\odot}}$. The LoS velocity dispersion inside $r_0$ is $81.2~{\rm km~s^{-1}}$.}
  \label{less_massive}
\end{figure}

The result is shown in Figure \ref{less_massive}. In comparison to the previous runs, the halo rotation due to the clumps is much weaker in this case. In the bottom panel, although a rotating signature can be seen weakly in $z\hspace{0.3em}\raisebox{0.4ex}{$<$}\hspace{-0.75em}\raisebox{-.7ex}{$\sim$}\hspace{0.3em}1~{\rm kpc}$ and $R=1 - 7~{\rm kpc}$, it is less significant than the previous cases and the orbital eccentricities are unchanged. Hence, clumps smaller than $1\times10^8~{\rm M_{\odot}}$ could not be expected to leave the kinematic imprint on halo objects.

It is important to remark that dynamical friction depends on clump mass. In the Chandrasekhar formula, the deceleration is proportional to the clump mass and the low-mass clumps slowly migrate into the galactic center \citep{c:43,bt:08}. Since the clump masses are five times smaller, the time scale of orbital shrinkage of the clumps is approximately five times longer than that in the fiducial case. Although such lower mass clumps have a longer time to interact with the halo objects, the imprinted rotation on the halo is much weaker. Accordingly, I see that the clump interaction with halo objects seems to strongly depend on the clump masses. 
 
\subsection{The case with mass loss}
\label{mass-loss}
\citet{n:96} has pointed out the possibility that giant clumps may be fragile objects if they suffer mass loss from outflow strong enough to blow gas out. Recently, \citet{gnj:11} and \citet{nsg:12} have actually observed such outflow from clumps in some galaxies, which exceeds escape velocities of the clumps. Furthermore, \citet{hkm:11} and \citet{g:12} performed numerical simulations that suggest that such superwinds can blow gas away from clumps and prevent the clumps from growing massive. As a result, the clumps become vulnerable to tidal disruption. These studies discussed that clumps would be short lived and disrupted in a few dynamical times of the clump rotation. On the other hand, \citet{fsg:11} and \citet{ggf:11} observed radial-age gradients of clumps in some galaxies. They argued that the age gradients could be indicative of the migration and longevity of the clumps. Moreover, \citet{eef:09} discussed that clump ages  are intrinsically closer to ages of bulge-like objects in galaxies, in cases where there are bulge-like objects. They discussed that giant clumps are long lived and can reach the galactic center by dynamical friction.

In cases where the superwinds are real and clumps are short lived, clumps may not be able to interact with halo objects. Accordingly, I perform a run assuming mass loss from the clumps in the calculation to study how the mass loss caused by the superwinds and/or tidal disruption affects the kinematic imprint on a halo. In this run, the masses of the clumps are initially $M_{cl}=5\times10^8~{\rm M_{\odot}}$. The clumps decrease their masses at a rate of $0.743~{\rm \%}$ per time step, while their orbits are time integrated. At this mass loss rate, the clumps lose 99 \% of their masses in the first $t=0.35~{\rm Gyr}$ (one orbital period at $10~{\rm kpc}$ from the center).\footnote{This rate corresponds to $-6.56~{\rm M_\odot~yr^{-1}}$ for $M_{\rm cl}=5\times10^8~{\rm M_{\odot}}$.} In this case, the clumps do not build up a bulge because of the mass loss. The initial condition have ten clumps and the other settings are the same as the fiducial model. 

\begin{figure}
  \includegraphics[width=\hsize]{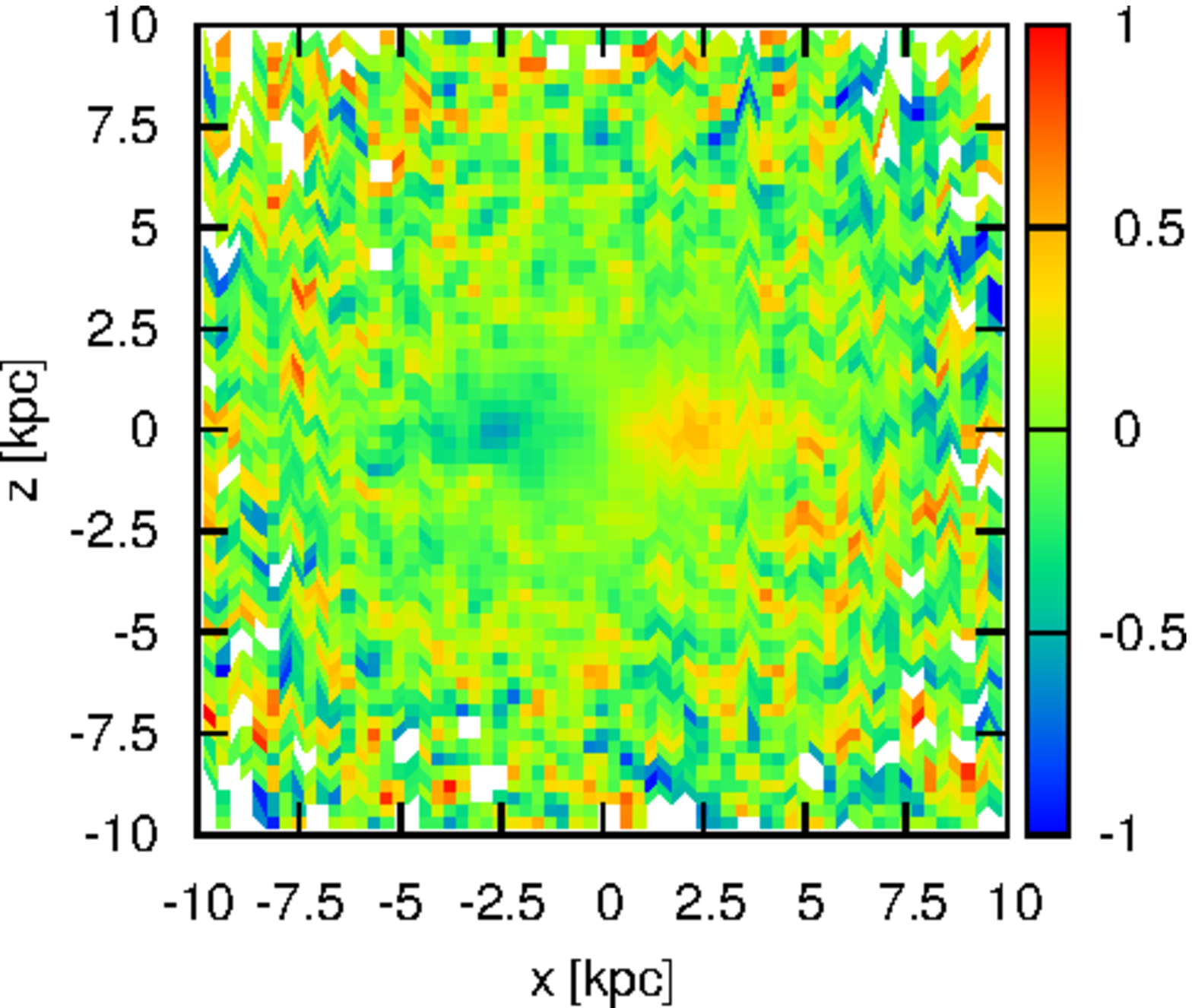}
  \includegraphics[width=\hsize]{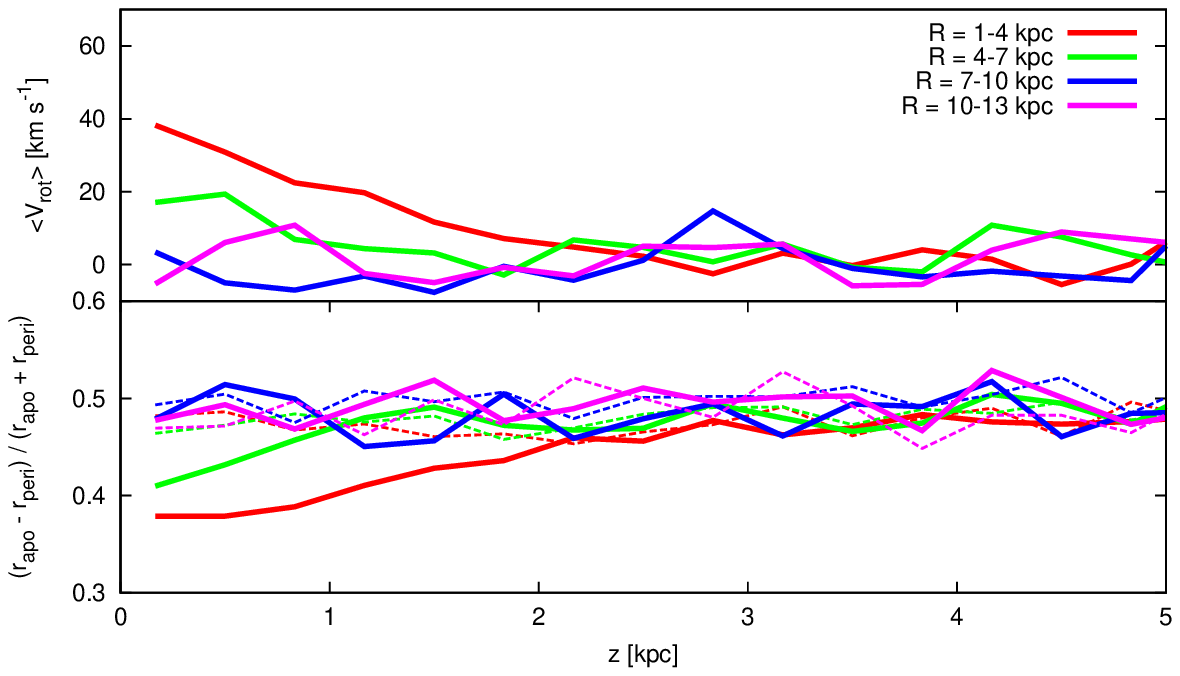}
  \caption{Same as Figure \ref{los} and Figure \ref{dist_z}, but the mass loss from the clumps is implemented. The values are normalized by the velocity dispersion inside $r_0=4~{\rm kpc}$, which is $75.0~{\rm km~s^{-1}}$.}
  \label{out}
\end{figure}
Figure \ref{out} displays a LoS velocity map, mean azimuthal velocities, and eccentricities in this case. Although halo rotation appears only weakly inside $R\hspace{0.3em}\raisebox{0.4ex}{$<$}\hspace{-0.75em}\raisebox{-.7ex}{$\sim$}\hspace{0.3em}4~{\rm kpc}$ and $z\hspace{0.3em}\raisebox{0.4ex}{$<$}\hspace{-0.75em}\raisebox{-.7ex}{$\sim$}\hspace{0.3em}2~{\rm kpc}$, the weak influence in the innermost region may be due to the initial set-up arbitrarily putting some clumps on the inner radii. Here, the mass loss time scale becomes relatively long compared to orbital time scale. As noted in \S\ref{lowmass}, since the deceleration by dynamical friction is proportional to the clump mass in the Chandrasekhar formula, the clumps losing their masses can keep their orbits for a long time. However, as shown here, the clumps subjected to the rapid mass loss do not seem to exercise gravity on the halo objects significantly. Thus, I suggest that the existence of the rapid mass loss from clumps could be a crucial factor of the kinematic imprint of clumpy disk formation, which is still under debate.

\section{Discussion}
\label{dis}
From the results above, I suggest that if giant clumps in a clumpy disk galaxy can retain their masses of $\hspace{0.3em}\raisebox{0.4ex}{$>$}\hspace{-0.75em}\raisebox{-.7ex}{$\sim$}\hspace{0.3em}5\times10^8~{\rm M_\odot}$ until reaching the galactic center, they could gravitationally interact with halo objects, such as old GCs and HSs, even in the time scale of dynamical friction, $t\simeq0.5~{\rm Gyr}$. Hence, clumpy disk formation can be expected to leave a kinematic imprint on the halo objects, which may still remain in a current galactic halo if the galaxy has not experienced violent disturbances, such as major mergers after the clumpy phase. In such a case, the influenced halos would show net rotation with vertical gradients of mean velocity and orbital eccentricity around a disk plane.

The calculations, however, seem to show that the significance of the kinematic influence strongly depends on the clump mass. If the clumps are less massive than $1\times10^8~{\rm M_\odot}$ or if they suffer violent mass loss, the kinematic state of the halo barely changes. This means that clamp mass (mass loss rate) is a key to whether or not we can deduce remnants of clumpy galaxies from halo kinematics in current disk galaxies. From the result of \S\ref{lowmass}, I estimate that the critical mass to exercise influence on a halo would be $M_{cl}\sim5\times10^8~{\rm M_\odot}$ if there is no mass loss. Present observations have observed clumps more massive than this criterion in clumpy galaxies. Therefore the existence of such massive clumps is not unrealistic. However, current high redshift observations are biased to relatively luminous galaxies, and such brighter galaxies are naively expected to host more massive clumps than fainter galaxies \citep[e.g.,][]{eer:07,eef:09,eem:09}. Hence, although the massive clumps can exist, their typical mass may be lower than the critical mass, and the clump masses could depend on the mass of the host galaxy.

In \S\ref{mass-loss}, I showed that the life time of clumps is also a key factor. Although I arbitrarily set the mass loss rate of the clumps (99 \% mass loss in $0.35~{\rm Gyr}$) in my calculation, simulations of \citet{g:12} have suggested much shorter life times of clumps, $\sim50~{\rm Myr}$ at $R\sim3~{\rm kpc}$. Therefore, I infer that the kinematic imprint would be even weaker than my calculation or could not be seen if such a high mass loss rate is real. However, it is still being actively debated whether the clumps are short or long lived, as I noted in \S\ref{mass-loss} \citep[see][]{gnj:11,nsg:12,fsg:11,ggf:11}. 

\begin{figure}
  \includegraphics[width=\hsize]{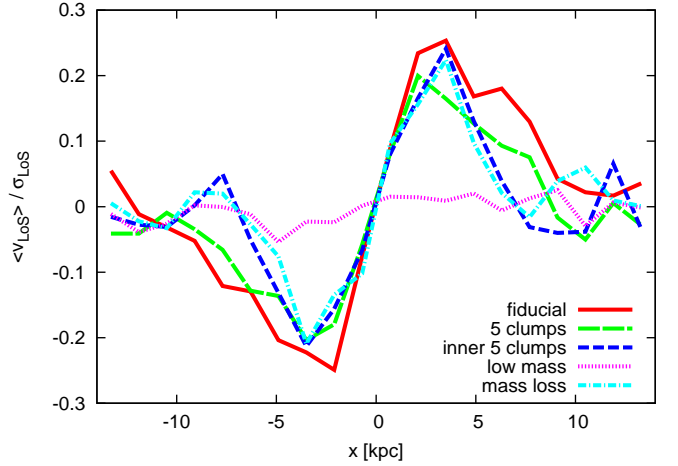}
  \caption{Profiles of mean LoS velocities over dispersions of the halo systems along the disk planes. The LoS dispersion is computed in each bin, \textit{not} in the central region. The red solid line indicates the results of \S\ref{fiducial}; the green long-dashed and the blue short-dashed lines indicate the cases of 5 clumps with separations of $2~{\rm kpc}$ and $1~{\rm kpc}$, respectively, in \S\ref{5clumps}; the pink dotted line indicates the result of \S\ref{lowmass}; the light blue dot-dashed lines indicates the results of \S\ref{mass-loss}.}
  \label{rotation_curve}
\end{figure}

\begin{figure}
  \includegraphics[width=\hsize]{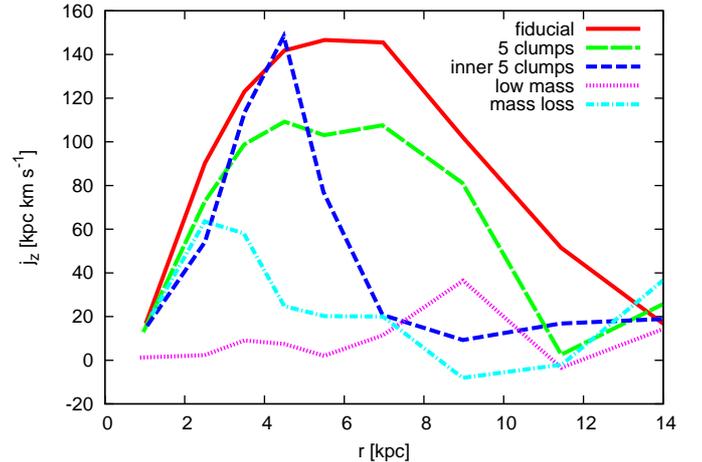}
  \caption{Spherically averaged specific angular momentum profiles of the halo systems. The line types are the same as in Figure \ref{rotation_curve}.}
  \label{SpeAng}
\end{figure}

Other dynamical processes may be able to provide a halo system with angular momentum. Numerical simulations by \citet{bfb:02,bbb:05} have demonstrated that metal-poor GCs can obtain a net rotation by galactic mergers. However, their results showed that the GC system rotates significantly in an outer halo region rather than in an inner region. Moreover, the direction of the rotation would not necessarily coincide with the disk rotation in the merger scenario. \citet{beb:11} conducted binary merger simulations with large samples, which also indicate that merger remnants generally have larger spin parameters in their outer regions than in their inner regions. In addition, \citet{djl:09} demonstrated that a merger remnant of non-rotating spherical galaxies can be a rotating elliptical. However, also in their simulation, an outer region of the remnant rather than an inner region, indicates significant rotation. In Figure \ref{rotation_curve}, I show ratios of LoS velocities to LoS velocity dispersions. The ratios decline sharply outside the disk radii, and the outer regions are not rotating. Thus the rotation signatures by clumpy disk formation are different from those induced by galactic mergers (compare to Figure 6 of \citet{djl:09}). Also in cosmological $N$-body simulations, DM halos generally have averaged rotations approximately proportional to $j\propto r^{1.1}$ \citep[e.g.,][]{bdk:01}. This indicates nearly constant or slightly increasing rotation velocities toward outer halos, where $j$ is specific angular momentum, although a stellar halo may not necessarily share the same kinematic state with a DM halo. In Figure \ref{SpeAng}, radial profiles of angular momenta in my calculations are shown. Obviously, the angular momentum profiles are not in monotonic fashion and show little rotation outside the disk radii. I therefore suggest that the disky halo rotation caused by clump motions could be unique and distinguishable from the cosmological merger-origin rotation. 

Having formed in a galactic disk, old stars may also have properties similar to the HSs influenced by clumps, even if the disk has not experienced a clumpy phase. These old stars can be heated up in the vertical direction by perturbations such as passages of spiral arms and giant molecular clouds. Eventually the old disk stars can have a thickened density distribution. However, the disky halo rotation by clumps is much slower than disk rotation. As a result, the rotating HSs could be distinguished kinematically (and also chemically) from the old disk stars.

Meanwhile, GCs associated with a galactic disk (disk GCs), which have been observed in some galaxies, also have a disky kinematic state similar to my result \citep{z:85,cbf:02,mhp:04,mg:04,mb:05}. However, the disk GCs are metal-rich and thought to have formed in a relatively late stage of galaxy evolution, although there seems to be a handful of old and metal-poor disk GCs \citep[e.g.,][]{dgv:99,dgv:03,mg:04}. Furthermore, their orbits almost coincide with disk rotations, indicating nearly circular rotation velocities and low orbital eccentricities, which are inconsistent with my results (Figure \ref{dist_z}: slow rotation and mild orbital circularization). Thus, I expect that the disk GCs are not indicative of the clump interaction and that clumpy disk formation does not explain the origin of the disk GCs.

However, origins of old GCs and HSs have not been elucidated yet and their initial states of distribution and kinematics are thus unclear \citep[see][]{bs:06}. Some studies have suggested that a part of GCs and HSs seems to have external origins from accreting dwarf galaxies \citep[e.g.,][]{cmw:98,ans:06,mhf:10}. Recently, \citet{tb:12} observed stellar halos of two Milky Way-like galaxies at a redshift of $z\sim1$, which seem to have structural properties similar to those of counterparts in the local universe. They suggest, however, that some accretion and/or star formation is still lasting after $z\sim1$. If it is the case, acquisition by a galaxy of GCs and HSs may postdate the clumpy disk formation and the halo objects originating from such late accretion can avoid the influence in the clumpy disk formation stage. 
 
What fraction of disk galaxies can have the imprints of past clumpy phases? If it is here assumed that a fraction, $A$, of current disk galaxies have experienced clumpy disk formation and that a fraction, $B$, of the clumpy galaxies have hosted giant clumps of $M_{cl}>5\times10^8~{\rm M_\odot}$, it is simply expected that approximately $A\times B$ of the current disk galaxies would show the imprinted signatures, even though the clumps have already disappeared to the current galaxies. However, it is still challenging to estimate the fractions of $A$ and $B$ in present observations that would be biased toward luminous galaxies. In this regard, cosmological simulations may be powerful tools to explore this issue, although current simulations are still lacking samples with high resolution \citep[e.g.,][]{mgc:12}.

Are there remnants of clumpy galaxies nearby? Some local disk galaxies have indeed been observed to have a net rotation in their old (metal-poor) GC system, such as the Large Magellanic Cloud \citep[e.g.,][]{fio:83}, M31 \citep[e.g.,][]{pbh:02}, NGC253 \citep{oms:04}, NGC524 \citep{bfb:04}, NGC2683 \citep{pfb:08}, NGC3115 \citep{kzs:02}.\footnote{In M33, GCs older than $1~{\rm Gyr}$ have been observed to have a significant rotation, but the number of observed samples older than several gigayears is only a handful \citep{cbf:02}.} However, because of their faintness, the number of GCs observed spectroscopically to measure their kinematics is still limited in extra-galaxies. Furthermore, it would be observationally difficult to pick out a sufficiently large number of GCs from their disk regions. In the Milky Way, there seems to be little net rotation in metal-poor GCs in any radial ranges (e.g., \citet{bs:06}\footnote{However, \citet{bs:06} have mentioned that a strong prograde rotation is seen in the most metal-poor GCs.}, although see \citet{z:85}). Meanwhile, rotation of HSs in the Milky Way is still controversial. Some observations have suggested a slow prograde rotation, $<50~{\rm km~s^{-1}}$, of the inner halo \citep{cb:00,kmh:07,cbl:07,slf:10}\footnote{\citet{cbl:07} did not reject the possibility of a non-rotating inner halo.}, but others have favored a non-rotating inner halo \citep{vpb:06,seb:09,bis:10}. Thus, present observations do not seem to enable us to identify galaxies that were once clumpy disks. A near-future astrometry satellite mission, \textit{Gaia}, is promising in this regard.

It should be noted that my calculations were performed with the simplistic toy models. In a real galaxy, disk formation and growth of DM halo follow the clumpy phase. A massive disk can deform a galactic potential into an axisymmetric (disky) one. Then, a distribution of halo objects would become aspherical, like Figure \ref{dist}, by the disk potential even if there were no clumps in a disk-formation stage. Moreover, using a numerical simulation  \citet{is:11} have demonstrated that a DM halo initially assumed to be an NFW profile is turned into a cored profile in a clumpy galaxy, though I assumed the rigid NFW halo potential. More sophisticated and realistic simulations should be performed in future works.

\section{Summary}
\label{sum}
If halo objects such as GCs and HSs have formed by internal processes of galaxy evolution, clumpy disk formation can be expected to have taken place in a pre-existing halo consisting of the old objects. In such a case, the clump orbital motions may be able to exercise influence on the kinematic state of the halo objects. Accordingly, I performed restricted $N$-body calculations and studied the kinematic imprint of the clumpy phase on a halo. I found that the halo objects can be expected to indicate slow rotation and orbital circularization around the galactic disk after the clumpy phase. I also found that postulated clump masses must be more massive than $5\times10^8~{\rm M_\odot}$ and five clumps are enough to leave the imprint.

\begin{acknowledgements}
The author appreciates the referee for constructive and fascinating suggestions and is grateful to Daisuke Kawata and Kohei Hattori for their helpful discussion, and to Naoteru Gouda and Keith Waddington for reading proofs. This research was partially supported by the Ministry of Education, Science, Sports and Culture, Grant-in-Aid for Scientific Research (A), No.23244034, 2011-2015. 
\end{acknowledgements}

\bibliographystyle{aa}

\end{document}